\documentclass[aps,prl,twocolumn,superscriptaddress,showpacs,preprintnumbers,amsmath,amssymb]{revtex4}

\usepackage{graphicx} 
\usepackage{dcolumn}  

\graphicspath{{figs/}}

\def\nbb{\mbox{$275$}}

\def\SIGMA{\mbox{$5.8$}}

\def\BR{
  \mbox{$(2.3^{+0.4+0.2}_{-0.5-0.3}
  )\times 10^{-6}$}}

\def\fb{\mbox{fb$^{-1}$}}
\def\bb{\mbox{$B\overline{B}$}}
\def\qq{\mbox{$q\overline{q}$}}
\def\mbc{\mbox{$M_{\rm bc}$}}
\def\de{\mbox{$\Delta E$}}
\def\acp{\mbox{${\cal A}_{CP}$}}

\def\bz{\mbox{$B^0$}}

\def\bp{\mbox{$B^+$}}

\def\dz{\mbox{$D^0$}}
\def\dzb{\mbox{$\overline{D}{}^0$}}

\def\kppim{\mbox{$K^+ \pi^-$}}

\def\pizpiz{\mbox{$\pi^0 \pi^0$}}

\def\dzbpip{\mbox{$\dzb(\to\kppim\pi^0)\pi^+$}}

\begin{document}

%

\title{\quad\\[0.5cm] \boldmath
  Observation of $\bz \to \pizpiz$}

\affiliation{Aomori University, Aomori}
\affiliation{Budker Institute of Nuclear Physics, Novosibirsk}
\affiliation{Chiba University, Chiba}
\affiliation{Chonnam National University, Kwangju}
\affiliation{Chuo University, Tokyo}
\affiliation{University of Cincinnati, Cincinnati, Ohio 45221}
\affiliation{University of Frankfurt, Frankfurt}
\affiliation{Gyeongsang National University, Chinju}
\affiliation{University of Hawaii, Honolulu, Hawaii 96822}
\affiliation{High Energy Accelerator Research Organization (KEK), Tsukuba}
\affiliation{Hiroshima Institute of Technology, Hiroshima}
\affiliation{Institute of High Energy Physics, Chinese Academy of Sciences, Beijing}
\affiliation{Institute of High Energy Physics, Vienna}
\affiliation{Institute for Theoretical and Experimental Physics, Moscow}
\affiliation{J. Stefan Institute, Ljubljana}
\affiliation{Kanagawa University, Yokohama}
\affiliation{Korea University, Seoul}
\affiliation{Kyoto University, Kyoto}
\affiliation{Kyungpook National University, Taegu}
\affiliation{Swiss Federal Institute of Technology of Lausanne, EPFL, Lausanne}
\affiliation{University of Ljubljana, Ljubljana}
\affiliation{University of Maribor, Maribor}
\affiliation{University of Melbourne, Victoria}
\affiliation{Nagoya University, Nagoya}
\affiliation{Nara Women's University, Nara}
\affiliation{National Central University, Chung-li}
\affiliation{National Kaohsiung Normal University, Kaohsiung}
\affiliation{National United University, Miao Li}
\affiliation{Department of Physics, National Taiwan University, Taipei}
\affiliation{H. Niewodniczanski Institute of Nuclear Physics, Krakow}
\affiliation{Nihon Dental College, Niigata}
\affiliation{Niigata University, Niigata}
\affiliation{Osaka City University, Osaka}
\affiliation{Osaka University, Osaka}
\affiliation{Panjab University, Chandigarh}
\affiliation{Peking University, Beijing}
\affiliation{Princeton University, Princeton, New Jersey 08545}
\affiliation{RIKEN BNL Research Center, Upton, New York 11973}
\affiliation{Saga University, Saga}
\affiliation{University of Science and Technology of China, Hefei}
\affiliation{Seoul National University, Seoul}
\affiliation{Sungkyunkwan University, Suwon}
\affiliation{University of Sydney, Sydney NSW}
\affiliation{Tata Institute of Fundamental Research, Bombay}
\affiliation{Toho University, Funabashi}
\affiliation{Tohoku Gakuin University, Tagajo}
\affiliation{Tohoku University, Sendai}
\affiliation{Department of Physics, University of Tokyo, Tokyo}
\affiliation{Tokyo Institute of Technology, Tokyo}
\affiliation{Tokyo Metropolitan University, Tokyo}
\affiliation{Tokyo University of Agriculture and Technology, Tokyo}
\affiliation{Toyama National College of Maritime Technology, Toyama}
\affiliation{University of Tsukuba, Tsukuba}
\affiliation{Utkal University, Bhubaneswer}
\affiliation{Virginia Polytechnic Institute and State University, Blacksburg, Virginia 24061}
\affiliation{Yonsei University, Seoul}
  \author{Y.~Chao}\affiliation{Department of Physics, National Taiwan University, Taipei} 
  \author{P.~Chang}\affiliation{Department of Physics, National Taiwan University, Taipei} 
  \author{K.~Abe}\affiliation{High Energy Accelerator Research Organization (KEK), Tsukuba} 
  \author{K.~Abe}\affiliation{Tohoku Gakuin University, Tagajo} 
  \author{H.~Aihara}\affiliation{Department of Physics, University of Tokyo, Tokyo} 
  \author{K.~Akai}\affiliation{High Energy Accelerator Research Organization (KEK), Tsukuba} 
  \author{M.~Akatsu}\affiliation{Nagoya University, Nagoya} 
  \author{M.~Akemoto}\affiliation{High Energy Accelerator Research Organization (KEK), Tsukuba} 
  \author{Y.~Asano}\affiliation{University of Tsukuba, Tsukuba} 
  \author{T.~Aushev}\affiliation{Institute for Theoretical and Experimental Physics, Moscow} 
% ...
\collaboration{The Belle Collaboration}

\noaffiliation


\tighten

\begin{abstract}
  We report the first observation of the decay $\bz\to\pizpiz$,
  using a 253 fb$^{-1}$ data sample collected at the
  $\Upsilon$(4S) resonance with the Belle detector at the
  KEKB $e^+e^-$ collider.
  The measured branching fraction is ${\cal B}(\bz\to\pizpiz) = \BR$,
  with a significance of $\SIGMA$ standard deviations including
  systematic uncertainties.
  We also make the first measurement of the direct $CP$ violating asymmetry 
  in this mode.
\end{abstract}


\maketitle


{\renewcommand{\thefootnote}{\fnsymbol{footnote}}}

Measurements of the mixing-induced $CP$ violation parameter
$\rm{sin} 2\phi_1$~\cite{phi1_belle,phi1_babar} at $B$ factories
are in good agreement with the Kobayashi-Maskawa (KM)
mechanism~\cite{km}. 
To confirm this theory, one now has to measure
the other two angles of the unitarity triangle, $\phi_2$ and $\phi_3$.
%
One technique for measuring $\phi_2$ is to
study~\cite{phi2_babar,phi2_belle} time dependent $CP$ asymmetries
in $B^0 \to \pi^+\pi^-$ decay, where we have recently
reported~\cite{apipi_belle} the observation of $CP$ violation and
evidence for direct $CP$ violation. The extraction of
$\phi_2$, however, is complicated by the presence of both tree and
penguin amplitudes, each with different weak phases. An isospin
analysis of the $\pi\pi$ system is necessary~\cite{isospin}, and
one essential ingredient is the branching
fraction for the decay $B^0 \to \pi^0\pi^0$.



QCD-based factorization predictions for ${\cal B}(B^0 \to
\pi^0\pi^0)$ are typically around or below $1 \times
10^{-6}$~\cite{pi0pi0_predictions}, but phenomenological models
incorporating large rescattering effects can accommodate larger
values~\cite{theory}. Evidence for $B^0 \to \pi^0\pi^0$
emerged~\cite{pi0pi0_BaBar,pi0pi0_Belle} at the $B$ factories a
year ago, with a combined value of $(1.9\pm0.5)\times
10^{-6}$ for the branching fraction~\cite{PDG}.
If such a high value persists, an isospin
analysis for $\phi_2$ extraction would become feasible in the near
future. To complete the program, one would need to measure both the
$B^0$ and $\overline{B}{}^0$ decay rates, i.e. direct $CP$ violation.

In this paper we report the first observation of the decay
$\bz\to\pizpiz$. We also make a first measurement of the direct $CP$
violating asymmetry in this mode. The results are based on a 253 $\fb$
($\nbb$ M $\bb$ pairs) dataset collected with the Belle detector
at the KEKB $e^+e^-$ asymmetric collider~\cite{kekb}. KEKB
operates at a center--of--mass (CM) energy of $\sqrt{s} = 10.58 \
{\rm GeV}$, corresponding to the mass of the $\Upsilon$(4S)
resonance. Throughout this paper, neutral and charged $B$ mesons
are assumed to be produced in equal amounts at the $\Upsilon$(4S),
and the inclusion of charge conjugate modes is implied, unless
otherwise specified.

The Belle detector is a large-solid-angle magnetic spectrometer
that consists of a silicon vertex detector (SVD), a 50-layer
central drift chamber (CDC), an array of aerogel threshold
\v{C}erenkov counters (ACC), a barrel-like arrangement of
time-of-flight scintillation counters (TOF), and an
electromagnetic calorimeter (ECL) comprised of CsI(Tl) crystals
located inside a superconducting solenoid coil that provides a
1.5~T magnetic field.  An iron flux-return located outside of the
coil is instrumented to detect $K_L^0$ mesons and to identify
muons (KLM).  The detector is described in detail
elsewhere~\cite{Belle}. Two different inner detector
configurations were used. For the first sample of 152 million
$\bb$ pairs (Set I), a 2.0 cm radius beampipe and a 3-layer
silicon vertex detector were used; for the latter  123 million
$\bb$ pairs (Set II), a 1.5 cm radius beampipe, a 4-layer silicon
detector and a small-cell inner drift chamber were
used~\cite{Ushiroda}.

Pairs of photons with invariant masses in the range $115 \ {\rm
MeV}/c^2 < m_{\gamma\gamma} < 152 \ {\rm MeV}/c^2$ are used to
form $\pi^0$ mesons; this corresponds to a window of $\pm
2.5\sigma$ about the nominal $\pi^0$ mass, where $\sigma$ denotes
the experimental resolution, approximately $8 \ {\rm MeV}/c^2$.
The measured energy of each photon in the laboratory frame is
required to be greater than $50 \ {\rm MeV}$ in the barrel region,
defined as $32^{\circ} < \theta_{\gamma} < 129^{\circ}$, and
greater than $100 \ {\rm MeV}$ in the end-cap regions, defined as
$17^{\circ} \le \theta_{\gamma} \le 32^{\circ}$ and $129^{\circ}
\le \theta_{\gamma} \le 150^{\circ}$, where $\theta_{\gamma}$
denotes the polar angle of the photon with respect to the positron beam
line. To further reduce the combinatorial background, $\pi^0$
candidates with small decay angles ($\cos\theta^* >0.95$) are
rejected, where $\theta^*$ is the angle between the $\pi^0$ boost
direction from the laboratory frame and one of its $\gamma$ daughters
in the $\pi^0$ rest frame.

Signal $B$ candidates are formed from pairs of $\pi^0$ mesons and
are identified by their beam energy constrained mass $\mbc =
\sqrt{E_{\rm beam}^{*2} - p_B^{*2}}$ and energy difference $\de =
E_B^* - E_{\rm beam}^*$, where $E_{\rm beam}^*$ denotes the beam
energy and $p_B^*$ and $E_B^*$ are the momentum and energy,
respectively, of the reconstructed $B$ meson, all evaluated in the
$e^+e^-$ CM frame. We require $\mbc > 5.2 \ {\rm GeV}/c^2$ and
$-0.3 \ {\rm GeV} < \de < 0.5 \ {\rm GeV}$. The signal
efficiency 
is estimated using
GEANT-based~\cite{geant} Monte Carlo (MC) simulations. 
The resolution for signal is approximately $3.6 \ {\rm
MeV}/c^2$ in $\mbc$. The distribution in $\de$ is asymmetric due to
energy leakage from the CsI(Tl) crystals. If it is parameterized by a
bifurcated Gaussian, the upper and lower resolutions are $46 \ { \rm MeV}$
and $122 \ { \rm MeV}$, respectively.

We consider background from other $B$ decays and from
$e^+e^-\to\qq$ ($q = u$, $d$, $s$, $c$) continuum processes. A
large generic MC sample shows that backgrounds from $b\to c$
decays are negligible. Among charmless $B$ decays, the only
significant background is $B^\pm \to \rho^\pm\pi^0$ with a
missing low momentum $\pi^\pm$. This background populates the negative
$\de$ region, and is taken into account in the signal extraction
described below.

The dominant background is due to continuum processes. We use
event topology to discriminate signal events from this $\qq$
background, and follow the continuum rejection technique
from our previous publication~\cite{pi0pi0_Belle}. We use
modified Fox-Wolfram moments~\cite{fw} where the particles in the
signal $B$ candidate (category $s$) and those in the rest of the event
 (category $o$)
are treated separately; we also use the missing momentum of the event as a
third category (category $m$). Some additional discrimination is achieved
by considering charged and neutral particles in the $o$ category
independently, and by taking the correlations of charges into
account. We combine 16 modified moments with the scalar sum of the
transverse momentum into a Fisher discriminant~\cite{fisher}, and
tune the coefficients to optimize the separation between signal and background.

The angle of the $B$-meson flight direction with respect to the
beam axis ($\theta_B$) provides additional discrimination. A
likelihood ratio ${\cal R}_s = {\cal L}_s / ({\cal L}_s + {\cal
L}_{q\overline{q}})$ is used as the discrimination variable, where
${\cal L}_s$ denotes the product of the individual Fisher and
$\theta_B$ likelihoods for the signal and ${\cal
L}_{q\overline{q}}$ is that for the $\qq$ background. The
likelihood functions are derived from MC for the signal and from
events in the $\mbc$ sideband region ($5.20 \ {\rm GeV}/c^2 < \mbc
< 5.26 \ {\rm GeV}/c^2$) for the $\qq$ background.

Additional discrimination between signal and background can be
achieved by using the Belle standard algorithm for $b$-flavor
tagging~\cite{phi1_belle,phi2_belle}, which is also needed for the direct
$CP$ violation measurement. The flavor tagging procedure yields two
outputs: $q = \pm 1$, indicating the flavor of the other $B$ in the event,
and $r$, which takes values between 0 and 1 and
is a measure of the confidence that the $q$ determination is
correct. Events with a high value of $r$ are considered
well-tagged and are therefore unlikely to have originated from
continuum processes. For example, an event that contains a high momentum lepton
($r$ close to unity) is more likely to be a $B \overline B$ event
so a looser $\mathcal{R}_s$ requirement can be applied. We find
that there is no strong correlation between $r$ and any of the
topological variables used above to separate signal from
continuum.

We divide the data into $r\ge 0.5$ and $r<0.5$ bins. The continuum
background is reduced by applying a selection requirement on
$\mathcal{R}_s$ for events in each $r$ region of Set I and Set II
according to the figure of merit (FOM). The FOM is defined as
$N_s^{\rm exp}/\sqrt{N_s^{\rm exp}+N_{BG}^{\rm exp}}$, where
$N_s^{\rm exp}$ and $N_{BG}^{\rm exp}$ denote the expected signal,
assuming the branching fraction ${\cal B} = 2 \times 10^{-6}$,
and background yields obtained from MC and sideband data, respectively.
A typical requirement suppresses 97\% of the continuum background while
retaining 53\% of the signal.

The signal yields are extracted by applying unbinned two-dimensional
maximum likelihood (ML) fits to the ($M_{\rm bc}$,
$\Delta E$) distributions of the $B$ and $\overline B$ samples.
The likelihood is defined as
\begin{eqnarray}
\mathcal{L} & = & {\rm exp}\; (-\sum_{s,k,j} N_{s,k,j}) 
\prod_i (\sum_{s,k,j} N_{s,k,j} {\mathcal P}_{s,k,j,i}) \;\;\;
\end{eqnarray}
where
\begin{eqnarray} \mathcal{P}_{s,k,j,i} & = &
P_{s,k,j}(M_{{\rm bc}i}, \Delta E_i),
\end{eqnarray}
and $s$ indicates Set I or Set II, $k$ distinguishes events in the
$r<0.5$ or $r\ge 0.5$ bins, $i$ is the identifier of the $i$-th
event, $P_{s,k,j}(M_{\rm bc}, \Delta E)$ are the two-dimensional
probability density functions (PDFs) in $M_{\rm bc}$ and $\Delta E$
for the signal and background components, $N_j$ is the number of events for the
category $j$, which corresponds to either signal, $q\bar{q}$
continuum, or background from $B^\pm \to \rho^\pm\pi^0$ decay.

\begin{figure}
\resizebox{!}{0.25\textwidth}{\includegraphics{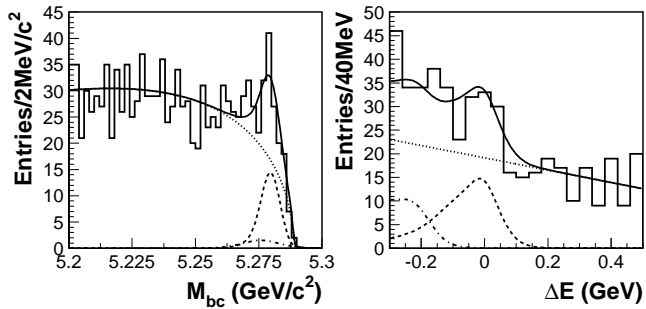}}
  \caption{
    \label{fig:fit_result}
    Result of the fit described in the text.
    (Left) $\mbc$ projection for events that satisfy
    $-0.2 \ {\rm GeV} < \de < 0.05 \ {\rm GeV}$;
    (right) $\de$ projection for events that satisfy
    $5.27 \ {\rm GeV}/c^2 < \mbc < 5.29 \ {\rm GeV}/c^2$.
    The solid lines indicate the sum of all components,
    and the dashed, dotted and dot-dashed lines
    represent the contributions from signal, continuum,
    and $B^+ \to \rho^+\pi^0$, respectively.
  }
\end{figure}

The PDFs for the signal and for $B^+ \to \rho^+\pi^0$ are taken
from smoothed two-dimensional histograms obtained from large MC
samples. For the signal PDF, discrepancies between the peak positions and
resolutions in data and MC are calibrated using $\dz\to\pizpiz$ and
$\bp\to\dzbpip$ decays. The difference is caused by imperfect simulation
of the $\pi^0$ energy resolution while the effect of the opening angle
distributions can be neglected. The invariant mass distribution for the $D^0$
is fitted with an empirical function for data and MC, and the
observed discrepancies in the peak position and width are converted to
the differences in the peak position and resolution for $\de$ in the
signal PDF. We require the $D^0$ decay products to lie in the same momentum
range as the $\pi^0$s from $B \to \pi^0\pi^0$. 
 To obtain the
two-dimensional PDF for the continuum background, we multiply the
PDF for $\de$, which is modeled with a linear function,
with the PDF for $\mbc$, for which we use the ARGUS
function~\cite{argus}. In the fit, the shapes of the signal and
$B^+ \to \rho^+\pi^0$ PDFs are fixed, with the normalization for
$B^+ \to \rho^+\pi^0$ floated; 
all other fit parameters are allowed to float.
 The fit results are shown in Fig.~\ref{fig:fit_result}.

The obtained signal yield is $81.8^{+15.5}_{-16.9}$ with a statistical
significance (${\cal S}$) of $6.1$, where ${\cal S}$ is defined as
${\cal S}=\sqrt{-2\ln({\cal L}_0/{\cal L}_{N_s})}$, and ${\cal
L}_0$ and ${\cal L}_{N_s}$ denote the maximum likelihoods of the
fits without and with the signal component, respectively. The relative
yields in sets I and II are consistent with the expectation based on their
relative luminosities.
We vary each calibration constant for the signal PDF by $\pm 1
\sigma$ and obtain systematic errors from the change in the signal
yield. Adding these errors in quadrature, 
the significance including systematic uncertainties
is reduced to $5.8 \sigma$, which corresponds to the first 
observation of $B^0\to \pi^0\pi^0$.

In order to obtain the branching fraction, we divide the signal
yield by the reconstruction efficiency, measured from MC to be
$12.9\%$, and by the number of $\bb$ pairs. 
We consider systematic errors in the
reconstruction efficiency due to possible differences between data
and MC. A $4.2\%$ systematic error is assigned for the uncertainty in the
efficiency for the track multiplicity requirement. This is determined by
varying the multiplicity distribution of signal MC.
 We assign a total error of $6\%$ due to $\pi^0$
reconstruction efficiency, measured by comparing the ratio of the
yields of the $\eta\to\pizpiz\pi^0$ and $\eta\to\gamma\gamma$
decays. The experimental errors on the branching fractions for
these decays~\cite{PDG} are included in this value. We check the
effect of the continuum suppression using a control sample of
$\bp\to\dzbpip$ decays; the ${\cal R}_s$ requirements has a similar
efficiency for the MC control sample and for signal MC. Comparing the
${\cal R}_s$ requirement on the control sample in data and MC,
a systematic error of $1.8\%$ is assigned.
We check for a possible pile-up background due to hadronic continuum
events that contain energy deposits from earlier QED interactions.
Such a background may peak in $\mbc$, however, the showers from
the QED interaction can be identified from timing information
recorded in the ECL. For Set II, it is
possible to remove these events using this information and determine
the change in event yield. We conservatively estimate a systematic
uncertainty of $10.3\%$ for this off-time QED background. 
Finally, we assign a systematic error of $1.1\%$ due to the
uncertainty in the number of $\bb$ pairs ($274.8 \pm 3.1$) $\times \
10^6$, and obtain a branching fraction of
\begin{displaymath}
{\cal B}(B^0\rightarrow\pi^0\pi^0)={\BR}.
\end{displaymath}
The result is stable under variations of the ${\cal R}_s$ cut.

\begin{figure}
\resizebox{!}{0.45\textwidth}{\includegraphics{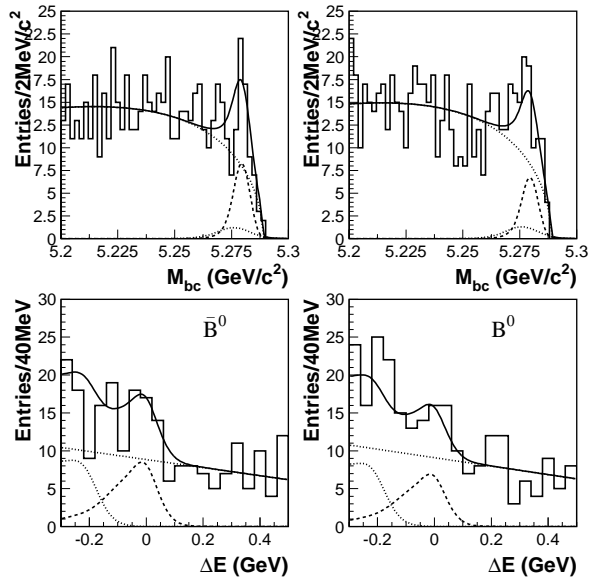}}
  \caption{
    \label{fig:tagged}
    $\mbc$ and $\de$ distributions with projections of the fit
    superimposed. The distributions are shown separately for
    events tagged as $\overline{B}{}^0$ (left) and ${B}^0$ (right).
  }
\end{figure}

Having observed a significant signal, we utilize the
$B^0$/$\overline{B}{}^0$ separation provided by the flavor tagging to
measure the $CP$ asymmetry. Equation (2) is replaced by
\begin{eqnarray} \mathcal{P}_{s,k,j} & = &\frac{1}{2}[1- q_i
\cdot \acp{'}_{l,j} ] P_{s,k,j}(M_{{\rm bc}i}, \Delta E_i),
\end{eqnarray}
where $q$ indicates the $B$ meson flavor, $B (q=+1)$ or
$\overline{B} (q=-1)$, $\acp{'}_{l,j}$ is the effective charge
asymmetry, where $\acp{'}_{l,j}=\acp{_j} (1-2\chi_d)(1-2 w_l)$.
Here $\chi_d=0.186\pm0.004$ \cite{PDG}
is the time-integrated mixing parameter and $w_l$ is the
wrong-tag fraction. For $\qq$ continuum, $\chi_d$ and $w_l$ are set to zero.
The $\pi^0\pi^0$ sample is divided into six $r$-bins,
and the $r$-dependent wrong-tag fractions, $w_l$ ($l=1,\ldots,6$),
are determined using a high statistics sample of self-tagged
$B^0 \to D^{(*)-} \pi^+, D^{*-} \rho^+$ and $D^{*-} \ell^+ \nu$
events and their charge conjugates \cite{tagging}.
The total number of signal events is fixed to the yield obtained from
the branching fraction measurement.
The relative fractions of signal events, $\qq$ and $\rho^\pm \pi^0$
background events in the different $r$ bins are also fixed.

Defining the direct $CP$ asymmetry as
\begin{eqnarray}
\acp \equiv \frac{N(\overline B \to \overline f)-N(B \to f)}
{N(\overline B \to \overline f)+N(B \to f)},
\end{eqnarray} the result is ${\cal A}_{CP} = 0.44^{+0.53}_{-0.52}\pm0.17$.
Systematic errors are estimated by varying the fitting parameters by
$\pm 1\sigma$. Including the result of a null asymmetry check with the
same analysis procedure for the $B \to D(K\pi\pi^0)\pi$ control sample,
the total systematic error is $\pm0.17$.
To illustrate this asymmetry, we show the results separately for
$B^0$ and $\overline{B}{}^0$ tags in Fig.~\ref{fig:tagged}.
While not significant, the method already gives constraints on
$\phi_2$~\cite{ckmfitter}.

Our results confirm the previous evidence~\cite{pi0pi0_BaBar,pi0pi0_Belle}
and establish the decay $B^0\to \pi^0\pi^0$.
Since the observed branching fraction is much larger than predictions
based on QCD factorization~\cite{pi0pi0_predictions}, recent
theoretical discussions have focused on the possibility of an enhanced
color-suppressed amplitude, together with a sizable strong phase~\cite{CGRS}.
Other color-suppressed modes such as ${B}^0\to {\overline D}{}^0\pi^0$
and $B^0\to \rho^0\pi^0$ have also
been measured~\cite{D0h0,rho0pi0} at rates considerably higher than
factorization predictions~\cite{theory,Barshay}.
In addition, the recent evidence for large
direct $CP$ violation in $B^0 \to \pi^+\pi^-$~\cite{apipi_belle} and
$B^0 \to K^+\pi^-$ modes~\cite{AcpKpi} disagrees with
QCD based factorization predictions. 
Some effect beyond factorization appears to
be present in charmless two-body $B$ decays. 

In conclusion, we have observed the $\bz\to\pizpiz$ decay mode
in a data sample of 275 million $\bb$ pairs with a branching
fraction significantly higher than factorization predictions. We
obtain $81.8^{+15.5}_{-16.9}$ signal events with a significance of
$5.8$ standard deviations ($\sigma$) including
systematic uncertainties. The branching fraction is measured to be
$\BR$. This result is consistent with, and supersedes, our
previous result. We have also made a first measurement of the
direct $CP$ violating asymmetry. The large branching fraction for
$\bz\to\pizpiz$, together with the measurements of its direct $CP$
violating asymmetry ${\cal A}_{CP}$, will allow a model-independent
extraction of the CKM angle $\phi_2$ from measurements
of the $B\to \pi\pi$ system in the near future.

We thank the KEKB group for the excellent operation of the
accelerator, the KEK Cryogenics group for the efficient operation
of the solenoid, and the KEK computer group and the NII for
valuable computing and Super-SINET network support.  We
acknowledge support from MEXT and JSPS (Japan); ARC and DEST
(Australia); NSFC (contract No.~10175071, China); DST (India); the
BK21 program of MOEHRD and the CHEP SRC program of KOSEF (Korea);
KBN (contract No.~2P03B 01324, Poland); MIST (Russia); MESS
(Slovenia); NSC and MOE (Taiwan); and DOE (USA).


\begin{thebibliography}{99}
\bibitem{phi1_belle}
  K.~Abe {\it et al.} (Belle Collaboration),
  Phys. Rev. D {\bf 66}, 071102(R) (2002).
\bibitem{phi1_babar}
  B.~Aubert {\it et al.} (BaBar Collaboration),
  Phys. Rev. Lett. {\bf 89}, 201802 (2002).
\bibitem{km}
  M.~Kobayashi and T.~Maskawa,
  Prog. Theor. Phys. {\bf 49}, 652 (1973).
\bibitem{phi2_babar}
  B.~Aubert {\it et al.} (BaBar Collaboration),
  Phys. Rev. Lett. {\bf 89}, 281802 (2002).
\bibitem{phi2_belle}
  K.~Abe {\it et al.} (Belle Collaboration),
  Phys. Rev. D {\bf 68}, 012001 (2003).
\bibitem{apipi_belle}
  K.~Abe {\it et al.} (Belle Collaboration),
  Phys.\ Rev.\ Lett.\  {\bf 93}, 021601 (2004).
\bibitem{isospin}
  M.~Gronau and D.~London,
  Phys. Rev. Lett. {\bf 65}, 3381 (1990);
  M.~Gronau, D.~London, N.~Sinha and R.~Sinha,
  Phys. Lett. B {\bf 514}, 315 (2001).
\bibitem{pi0pi0_predictions}
  M.~Beneke and M.~Neubert, Nucl.\ Phys.\ B {\bf 675}, 333 (2003);
  Y.-Y.~Keum and A.I.~Sanda, Phys. Rev. D {\bf 67}, 054009 (2003). 
\bibitem{theory}
  W.S.~Hou and K.C.~Yang, Phys.\ Rev.\ Lett.\  {\bf 84}, 4806 (2000);
  C.K.~Chua, W.S.~Hou and K.C.~Yang,
  Mod.\ Phys.\ Lett.\ A {\bf 18}, 1763 (2003);
  A.J.~Buras {\it et al.}, Phys. Rev. Lett. 92, 101804 (2004).
\bibitem{pi0pi0_BaBar}
  B.~Aubert {\it et al.} (BaBar Collaboration),
  Phys.\ Rev.\ Lett.\  {\bf 91}, 241801 (2003).
\bibitem{pi0pi0_Belle}
  S. H. Lee, K. Suzuki {\it et al.} (Belle Collaboration),
  Phys.\ Rev.\ Lett.\  {\bf 91}, 261801 (2003).
\bibitem{kekb}
  S.~Kurokawa and E.~Kikutani,
  Nucl. Inst. and Meth. A {\bf 499}, 1 (2003).
\bibitem{Belle}
  A.~Abashian {\it et al.} (Belle Collaboration),
  Nucl. Inst. and Meth. A {\bf 479}, 117 (2002).
\bibitem{Ushiroda}
  Y. Ushiroda (Belle SVD2 Group),
  Nucl. Inst. and Meth. A {\bf 511}, 6 (2003).
\bibitem{geant}
  R.~Brun {\it et al.},
  GEANT 3.21, CERN Report No. DD/EE/84-1 (1987).
\bibitem{fw}
  G.~Fox and S.~Wolfram,
  Phys. Rev. Lett. {\bf 41}, 1581 (1978).
\bibitem{fisher}
  R.A.~Fisher,
  Annals of Eugenics {\bf 7}, 179 (1936).
\bibitem{argus}
  H.~Albrecht {\it et al.} (ARGUS Collaboration),
  Phys. Lett. B {\bf 241}, 278 (1990).
 \bibitem{PDG}
  Particle Data Group, S. Eidelman {\it et al.},
  Phys. Lett. B {\bf 592}, 1 (2004).
\bibitem{tagging}
  H. Kakuno {\it et al.}, 
  Nucl. Inst. and Meth. A {\bf 533}, 516 (2004).
\bibitem{ckmfitter}
  J. Charles {\it et al.}, 
  hep-ph/0406184.
\bibitem{CGRS}
  C.W.~Chiang, M.~Gronau, J.L.~Rosner and D.A.~Suprun,
  Phys. Rev. D {\bf 70}, 034020 (2004).
\bibitem{D0h0}
  K.~Abe {\it et al.}  (BELLE Collaboration),
  BELLE-CONF-0416, hep-ex/0409004 (2004);
  B.~Aubert {\it et al.} (BaBar Collaboration),
  Phys.\ Rev.\ D {\bf 69}, 032004 (2004);
  T.E.~Coan {\it et al.}  (CLEO Collaboration),
  {\it ibid}. {\bf 88}, 062001 (2002).
\bibitem{rho0pi0}
  J.~Dragic {\it et al.} (Belle Collaboration),
  Phys. Rev. Lett. {\bf 93}, 131802 (2004).
\bibitem{Barshay}
  S.~Barshay, L.M.~Sehgal and J.~van Leusen,
  Phys.\ Lett.\ B {\bf 591}, 97 (2004).
\bibitem{AcpKpi}
  B.~Aubert {\it et al.} (BaBar Collaboration),
  Phys.\ Rev.\ Lett.\  {\bf 93}, 131801 (2004).
  Y.~Chao {\it et al.} (Belle Collaboration),
  Phys.\ Rev.\ Lett.\  {\bf 93}, 191802 (2004).
\end{thebibliography}
\end{document}